\documentclass[conference]{IEEEtran}
\IEEEoverridecommandlockouts
\usepackage{multirow}
\usepackage{multicol}
\usepackage[utf8]{inputenc} 
\usepackage{kotex} 
\usepackage{subcaption}
\usepackage{hyperref}
\usepackage{cite}
\usepackage{amsmath,amssymb,amsfonts}
\usepackage{algorithm}
\usepackage{algorithmicx}
\usepackage{algpseudocode}
\usepackage{graphicx}
\usepackage{textcomp}
\usepackage{xcolor}
\usepackage{xspace}
\usepackage{soul}
\usepackage{kotex}

\usepackage{booktabs} 
\usepackage{hyperref}
\definecolor{linkcolour}{rgb}{0,0.2,0.6}
\definecolor{xgreen}{rgb}{0.2,0.6,0.0}
\definecolor{xred}{rgb}{0.7,0.1,0.0}
\hypersetup{colorlinks,breaklinks,citecolor=red!50, urlcolor=linkcolour, linkcolor=xgreen}

\usepackage{amsmath}
\usepackage{color}
\usepackage{array}
\usepackage{stfloats}
\usepackage{amsthm} 
\usepackage{amssymb}
\usepackage{float}
\usepackage{nccmath}
\usepackage{graphicx}
\usepackage{setspace}
\usepackage[normalem]{ulem}
\usepackage{booktabs}
\usepackage{subcaption}
\usepackage{mwe}
\usepackage{cite}
\usepackage{float}
\usepackage{physics}
\usepackage{graphicx}
\usepackage[export]{adjustbox}
\usepackage{amsmath,amssymb,amsfonts}
\usepackage{graphicx}
\usepackage{textcomp}
\usepackage{xcolor}
\usepackage[utf8]{inputenc} 
\usepackage{kotex} 

\usepackage{algorithm} 
\usepackage{algpseudocode}
\usepackage{grffile}

\usepackage[normalem]{ulem}

\usepackage{relsize}
\usepackage{hyperref}

\begin{document}

\title{Diffusion-based Quantum Error Mitigation using Stochastic Differential Equation}

\author{\IEEEauthorblockN{
    Joo Yong Shim and Joongheon Kim
}
\IEEEauthorblockA{
Department of Electrical and Computer Engineering, Korea University, Seoul, Republic of Korea
\\
E-mails: \texttt{shimjoo@korea.ac.kr}, 
\texttt{joongheon@korea.ac.kr}
}
}

\maketitle

\begin{abstract}
Unlike closed systems, where the total energy and information are conserved within the system, open systems interact with the external environment which often leads to complex behaviors not seen in closed systems. The random fluctuations that arise due to the interaction with the external environment cause noise affecting the states of the quantum system, resulting in system errors. To effectively concern quantum error in open quantum systems, this paper introduces a novel approach to mitigate errors using diffusion models. This approach can be realized by noise occurrence formulation during the state evolution as forward-backward stochastic differential equations (FBSDE) and adapting the score-based generative model (SGM) to denoise errors in quantum states. 
\end{abstract}

\section{Introduction}
Quantum computing is anticipated to complement classical algorithms in various industry sectors that require large-scale computations or demand high-speed operations \cite{10232949,10506692,nn23baek,iotj23park,apwcs21kim,10501929}. Discussions of quantum computing in an ideal scenario often overlook the inherent challenges associated with the practical implementation of quantum technologies. Quantum computers are inherently prone to errors caused by interactions between their qubits and the environment~\cite{breuer2002theory}. The interaction of quantum systems with their environment causes significant noises, dramatically affecting their operational efficiency and accuracy. This noise arises from various sources, including quantum decoherence and quantum interference, posing a substantial barrier to the reliability of quantum computations. 

For a long period, several types of research on quantum error mitigation (QEM)~\cite{wang2022quantumnat}, ranging from fundamental limits and theoretical frameworks to practical simulation tools for QEM, have been proposed such as zero-noise extrapolation which uses data at different noise levels to extrapolate the zero-noise limit~\cite{li2017efficient}. Probabilistic error cancellation, which models the actual quantum operations as combinations of ideal and erroneous operations, compensating for errors by sampling different operations based on their likelihood of occurrence~\cite{endo2018practical}.
The measurement error mitigation addresses errors in the measurement phase, which can distort the final results of quantum computations~\cite{geller2020efficient}.
This paper excludes state preparation and measurement (SPAM) errors, which occur at the beginning and end of actual operations, however focuses on mitigating gate noise occurring in open quantum systems. 

For the practical applications of quantum computing in real-world scenarios, several factors must be considered, in particular, \textit{i)} computational overhead~\cite{takagi2022fundamental}, \textit{ii)} accuracy, and \textit{iii)} generality (generally applicable)~\cite{quantumnas}. Each of these factors plays a crucial role in transitioning quantum computing from theory to practical, scalable applications. Quantum error correction and mitigation techniques typically require additional qubits and computational steps, increasing complexity and overhead. For example, implementing quantum error correction methods often requires a significant number of auxiliary qubits for encoding and measurement. In particular, this overhead can be a challenge, especially in the current noisy intermediate scale quantum (NISQ) era~\cite{bharti2022noisy}, where the limited number of qubits available in modern quantum computers~\cite{IBM2021Quantum}. The other important consideration is accuracy in quantum computing, which is directly impacted by the effectiveness of the error mitigation strategy. QEM aims to preserve the fidelity of quantum states by reducing or correcting errors as they occur, thereby increasing the accuracy of the computation. However, the accuracy of these techniques depends on their ability to correctly identify and eliminate errors without additional error occurrence during the process. This is challenging in practice due to imperfect gate operations, finite coherence times, and the probabilistic nature of quantum measurements. Finally, QEM techniques should be broadly applicable across different quantum computing platforms and algorithms in an ideal manner. However, many approaches can be applied only to the specific types of errors or system architectures. The generality of a technique also depends on how easily it can be integrated with various quantum computing tasks and algorithms. The corresponding techniques that require the extensive modification of the quantum circuit or are only compatible with specific types of quantum gates may be less useful. Developing universally applicable QEM strategies that can adapt to different noise environments and error types is a key area of research, aiming to make quantum computing more robust and flexible.

Accordingly, this paper presents a novel QEM approach in open systems utilizing diffusion-based models. Diffusion models, which are predominantly used in generative modeling for high-fidelity image generation, audio synthesis, text generation, and molecular design, leverage the concept of simulating a Markov chain. This simulation involves modeling the gradual transition of data from a complex distribution to a tractable one (typically Gaussian) and subsequently learning to reverse this process. Seminal works like the introduction to denoising diffusion probabilistic models (DDPMs) by \cite{ho2020denoising} and improvements by \cite{nichol2021improved} have reduced sampling times and enhanced sample quality. Additionally, the continuous-time framework proposed in score-based generative modeling through stochastic differential equations \cite{song2020score} provides a robust theoretical foundation, further enhancing efficiency in diffusion models.

In the context of QEM, this work adapts these principles to address noise in quantum systems by formulating the noisy gate dynamics into linear stochastic differential equations (SDE), thus framing the problem as a forward-backward SDE (FBSDE) challenge. This method deviates from traditional approaches that rely on complex equations or additional quantum gates for noise mitigation. Instead, it simplifies the master equation into an SDE format, which enables simpler computational models using state vectors, thereby reducing system complexity \cite{bassi2003stochastic}. A score-based generative diffusion model is employed to analyze and mitigate noise patterns predicted by the linear SDEs. 

\section{Modeling Open Quantum Systems}
The fundamental understanding of quantum systems and their dynamics often begins with the Shr\"{o}dinger equation, which describes the time evolution of a pure state of closed quantum systems. It can be fundamentally expressed as follows,

\begin{equation}
\frac{d}{dt} \ket{\psi_t} = -\frac{i}{\hbar} \ket{\psi_t}
\end{equation}
where $\ket{\psi_t}$ represents the state of the system at time $t$, $\hbar$ is the Planck's constant, and $i$ is the imaginary unit, respectively.

To extend the Shr\"{o}dinger equation to explain quantum systems that include mixed states (but still assume closed systems without environmental interaction),  the concept of density matrices is essentially required. A density matrix, denoted by $\rho=\sum_i p_i \ketbra{\psi_i}$, represents the mixed state of a quantum system in a set of states ${\ket{\psi_i}}$ with probabilities $p_i$. The von Neumann equation gives the evolution of a mixed state, as follows,

\begin{equation}
\label{eq:vonNeumannEq}
\frac{d}{dt}\rho_t = -\frac{i}{\hbar} [H_t, \rho_t]
\end{equation}
where $H$ is the Hamiltonian operator. In addition, $[A , B]$ denotes the commutator, defined as $AB - BA$. 
In open quantum systems, interaction with the environment leads to non-unitary evolution. This cannot be described solely by the Schr\"{o}dinger equation or von Neumann equation. Thus, the additional term $\mathcal{L}(\rho_t)$ is used, as follows,
\begin{equation}
\frac{d}{dt} \rho_t = -\frac{i}{\hbar}[H_t,\rho_t] + \mathcal{L}(\rho_t)
\end{equation}
and this can be also described as follows for more explicitly,
\begin{equation}
\label{eq:Lindbald}
\frac{d}{dt} \rho_t = -\frac{i}{\hbar}[H_t,\rho_t] + \gamma \sum_n \left[ L_n \rho L_n^\dagger - \frac{1}{2} \{L_n^\dagger L_n, \rho_t\} \right] 
\end{equation}
where the first term $-\frac{i}{\hbar}[H_t, \rho_t]$ is the Hamiltonian part, which describes the unitary evolution. In addition, the second term $\mathcal{L}(\rho_t)$ expresses the dissipating effects on the open system. Lastly, the $L_n$ usually represents jump operators and the $\sum_n L_n \rho L_n^\dagger - \frac{1}{2} \{L_n^\dagger L_n, \rho_t\}$ represents the effect caused by the interaction to the external environment, adding probability to certain states. The Lindblad equation~\cite{breuer2002theory} thus derived is the most general form for the generator of quantum dynamics for Markovian systems. This formulation allows to explain a broad range of physical phenomena, including relaxation and decoherence, to be modeled accurately.

The dissipating term $\mathcal{L}(\rho_t)$ can be defined by the summation of the effects of the device depolarization and the relaxation process. The depolarization, which arises due to the imperfections of the device, represents a process where a qubit gradually loses its purity and moves towards a maximally mixed state, irrespective of its initial state~\cite{di2023noisy}. The Lindblad operators for depolarization in a single-qubit system are defined by the Pauli matrices $\sigma_x, \sigma_y, \sigma_z$ as follows, 

\begin{multline}
\mathcal{L}_{\text{depolarization}}(\rho_t)= \\ \sum_{k \in \{x, y, z\}} \gamma_{d,k} \left( \sigma_k \rho \sigma_k - \frac{1}{2} \{\sigma_k^2, \rho\} \right)
\end{multline}
where $\gamma_{d,k}$ is the rate of depolarization over axes $x,y,z$. In addition, $\rho$ is the density matrix of the system and $\sigma_k$ are the Pauli matrices. Because $\sigma_k^2 = I$ holds for all Pauli matrices, it can be simplified as follows,
\begin{equation}
\mathcal{L}_{\text{depolarization}}(\rho_t)= \sum_{k \in \{x, y, z\}} \gamma_{d,k} \left( \sigma_k \rho_t \sigma_k - \frac{1}{2} \rho_t \right).
\end{equation}

Relaxation error results from interactions between quantum bits (qubits) and their environment. This interaction leads to an energy exchange that typically pushes the qubit towards a lower energy state, known as the ground state $|0\rangle$. The primary effects of relaxation noise include amplitude and phase damping~\cite{di2023noisy}. Amplitude damping \cite{benenti2019principles} is one significant effect of relaxation where a qubit loses energy to its surroundings and gradually transitions from a higher energy state (like the excited state) $|1\rangle$ to the ground state $|0\rangle$. The amplitude damping is characterized by the Lindblad operator $\sigma^-$ (the lowering operator) modeled as:
\begin{equation}
\mathcal{L}_{\text{amp}}(\rho_t) = \gamma_a \left( \sigma^- \rho_t \sigma^+ - \frac{1}{2} \left\{ \sigma^+ \sigma^-, \rho_t \right\} \right),
\end{equation}
where $\gamma_a$ is the rate of amplitude damping and $\rho$ is the density matrix of the system. Moreover, $\sigma^- = (X - iY)/2$ and $\sigma^+ = (X + iY)/2$ are the lowering and raising operators, respectively. Lastly, $\{A, B\}$ denotes the anti-commutator, defined as $AB + BA$.

Besides energy loss, relaxation can also cause dephasing, where the coherence between quantum states (the phase relationships) is lost. This affects the off-diagonal elements of the qubit's density matrix, leading to a loss of quantum information that is not necessarily related to energy exchange. Dephasing is characterized by the operator $\sigma_z$ (the Pauli Z operator), affecting the coherence between the computational basis states without energy exchange, modeled as follows,
\begin{equation}
\mathcal{L}_{\text{phase}}(\rho_t) = \gamma_p \left( \sigma_z \rho_t \sigma_z - \rho_t \right),
\end{equation}
where $ \gamma_p $ is the dephasing rate.

Combining these two effects, the total relaxation term that describes both amplitude damping and dephasing in a single qubit system can be expressed as follows,
\begin{equation}
\mathcal{L}_{\text{relaxation}}(\rho_t) = \mathcal{L}_{\text{amp}}(\rho_t) + \mathcal{L}_{\text{phase}}(\rho_t),
\end{equation}
and this can be as follows for more explicitly,
\begin{multline}
\mathcal{L}_{\text{relaxation}}(\rho_t) = \gamma_a \left( \sigma^- \rho_t \sigma^+ - \frac{1}{2} \left\{ \sigma^+ \sigma^-, \rho_t \right\} \right) + \\ \gamma_p \left( \sigma_z \rho_t \sigma_z - \rho_t \right).
\end{multline}

In practical terms, relaxation noise is a critical factor to consider in quantum computing and other quantum technologies because it can degrade the performance of quantum algorithms, leading to errors in quantum computations. 

The combined Lindblad term for both depolarization and relaxation can be as follows, 
\begin{equation}
\mathcal{L}(\rho_t) = \mathcal{L}_{\text{depolarization}}(\rho) + \mathcal{L}_{\text{relaxation}}(\rho),
\end{equation}
and finally, the dissipator term can be obtained by reformulating~\cite{manzano2020short} using Lindblad operators $L_n$, i.e., 
\begin{equation}
\mathcal{L}(\rho_t) = \gamma \sum_n \left[ L_n \rho L_n^\dagger - \frac{1}{2} \{L_n^\dagger L_n, \rho_t\} \right].
\end{equation}

\section{Forward-Backward Stochastic Differential Equation (FBSDE) Approach}\label{sec:SDE}
This paper proposes an FBSDE-based noise mitigation method, where the FBSDE is a class of stochastic processes consisting of two linked differential equations, i.e., one evolving forward in time and the other backward. Combining with score-based generative modeling (SGM)~\cite{song2020score}, this work theoretically shows the noise mitigation problem can be framed within the mathematical structure of FBSDE and solved by SGM. During this computation setup, the forward equation might model the progressive addition of noise to the data (thus simulating the forward evolution of the noise process), and the backward equation would represent the learned denoising path (the reverse-time generative model). This could provide a robust mathematical framework for understanding and analyzing the dynamics of diffusion models, potentially enhancing effectiveness by exploiting the properties of FBSDE.

\subsection{Problem Formulation in SDE}
Let's consider \eqref{eq:Lindbald}, i.e., the Lindblad Master Equation, which was previously defined as follows,
\begin{multline}
\frac{d}{dt} \rho_t = -\frac{i}{\hbar}[H_t,\rho_t] + \\ \gamma \sum_n \left[ L_n \rho L_n^\dagger - \frac{1}{2} \{L_n^\dagger L_n, \rho_t\} \right]. 
\end{multline}

Unraveling the defined density matrix into the equation for the state vector, it can be represented in terms of linear SDE~\cite{bassi2003stochastic}, as follows,
\begin{multline}
\label{eq:LindbladlinearSDE}
{d|\psi_t\rangle} = \left[-\frac{i}{\hbar} H \,{dt} + \right.\\ \left.\sum_n \left[ i \sqrt{\gamma}L \,{dW_t} - \frac{\gamma}{2} L^\dagger L  \,{dt} \right] \right]|\psi_t\rangle.
\end{multline}

\subsection{Score-based Generative Model (SGM)}
The proposed SGM is an advanced method in the field of machine learning for generating new data samples that are statistically similar to a given dataset. SGM is based on the concept of gradually adding noise to data until a simple, known distribution is reached (often a Gaussian distribution). The `score' in SGM modeling refers to the gradient of the log probability density of the data concerning the data itself. The key insight of SGM is that this score can guide the process of generating new data samples by indicating how to denoise or reverse the noise added to the data. Its process can be described in three steps, as follows.
\begin{enumerate}
    \item \textit{Adding Noise:} A series of transformations is performed, gradually adding noise to the data, moving from a complex data distribution to a simpler, noise-dominated distribution. This process is called the diffusion process, typically defined through an SDE.
    
    \item \textit{Training the Model:} The neural network is trained to estimate the score, which quantifies the direction and rate at which the probability density of the data increases. This training involves learning to predict the gradient of the log density at different noise levels.

    \item \textit{Generating Samples:} To generate new samples, the model starts with samples from the simple noise distribution (like Gaussian) and then applies a reverse process, guided by the learned scores, to progressively remove the noise. This reverse process can also be modeled by a reverse SDE.

\end{enumerate}
It can be noted that SGM can be described in the FBSDE problem. Then, the diffusion process can be formulated in the form of forward SDE, as follows,
        \begin{equation}
        \label{eq:fowardSDEEq}
        {dX_t} = f(X_t,t)\,{dt} + g(t)\,dW_t,
        \end{equation}
where $W$ represents a standard Wiener process based on Brownian motion, $f(X_t,t)$ is the drift coefficient, and $g(t)$ is the diffusion coefficient. If $f(X_t,t)$ and $g(t)$ satisfy Lipschitz conditions (i.e. if their derivatives are bounded), 
\begin{align}
|\nabla_x f(x, t)| &\leq \epsilon_1, \\
|\nabla_t g(t)| & \leq \epsilon_2,  
\end{align}
then, a unique strong solution can be derived~\cite{oksendal2003stochastic}.

If the prior is known (i.e., predefined distribution), one can sample $X_T$ from the prior sample. If the prior is predefined as Gaussian, $X_T$ will be the sample from the Gaussian distribution. The reverse process of diffusion can also be given by the backward 
 SDE~\cite{anderson1982reverse}, as follows,

\begin{multline}
\label{eq:backwardSDEEq}
{dX_t} = \left[f(X_t,t) - g^2(t)\nabla_x\log{p_t(X_t)} \right]\,dt + \\ g(t)\,{d\bar{W_t}}, 
\end{multline}
where $\bar{W_t}$ is the time reversal of the Wiener process in (\ref{eq:fowardSDEEq}). While the forward process gradually adds noise in very small increments as $t$ increases, the reverse formulation is traced as $t$ decreases in very small increments. If the score $\nabla_x\log{p_t(X_t)}$ from the marginal distribution $p_t(X_t)$ can be determined, then $X_0$ can be obtained from a random noise sample $X_T$.

\subsection{SGM for QEM}
In QEM, adapting SGM would involve conceptualizing quantum states and operations in a probabilistic framework similar to how data distributions are treated in classical SGM. The primary steps include the following items.
\begin{enumerate}
    \item \textit{Modeling the Noise in Quantum States:} Similar to the \textit{Adding Noise} process in SGM, one could model the evolution of quantum decoherence and operational errors as a stochastic process that transforms an ideal quantum state into a noisy state. This could also be described using forward SDE.
    
    \item \textit{Training the Model for Score Estimation:} The score in SGM corresponds to the gradient of the log probability density. For quantum states, this could translate to gradients of quantum state probabilities or amplitudes, providing a direction in which to adjust quantum states to mitigate errors. Learning these gradients would involve training neural networks capable of estimating these derivatives from noisy quantum states.

    \item \textit{Denoising Quantum States:} Using the learned scores, one could then attempt to reverse the noise process. This involves applying a series of quantum operations that gradually `denoise' the quantum error, moving it closer to its intended noise-free state. 
\end{enumerate}

The QEM problem can be defined as an FBSDE problem. First, it can reformulate the (\ref{eq:LindbladlinearSDE}) into the form of (\ref{eq:fowardSDEEq}), as follows, 

\begin{equation}
\label{eq:noisemodelingSDE}
{d|\psi_t\rangle} = f(\psi_t, t) \, {dt} + g(\psi_t, t) \, {dW_t}. 
\end{equation}
where the drift and diffusion terms can be expressed as, 
\begin{align}
f(\psi_t, t) &= -\frac{i}{\hbar} H|\psi_t\rangle  - \sum_n \frac{\gamma}{2} L_n^\dagger L_n|\psi_t\rangle, \\
g(\psi_t,t) &= \sum_n i \sqrt{\gamma}L_n|\psi_t,\rangle, 
\end{align}
then, by applying \eqref{eq:backwardSDEEq}, the denoising process of the state of noisy gates can be derived as follows,
\begin{multline}
    {d|\psi_t\rangle} = \left[f(\psi_t, t)\, - g^2(\psi_t,t)\nabla_\psi\log{p_t(|\psi_T\rangle)} \right]\,dt + \\ g(\psi_t,t) \,{d\bar{W_t}}.
\end{multline}

\subsection{Solving the FBSDE using SGM} 
This section introduces the process to solve the FBSDE problem using SGM. The first step, which is \textit{Adding Noise} in classical SGM and \textit{Modeling the Noise in Quantum States} in the proposed method is conducted in previous subsections, where \eqref{eq:fowardSDEEq} and \eqref{eq:noisemodelingSDE} are obtained as the solutions. After this step, in order to solve the FBSDE, prior distribution $p_T(\cdot)$ and the score function $\nabla \log p_t(\cdot)$ should be given. Then, the $p_T(x)$ is usually assumed to be Gaussian, which is fully tractable using the diffusion process. In the QEM problem, $p_T(|\psi\rangle)$ is the noisy state. To estimate the score $\nabla \log p_t(\cdot)$, time-dependent neural network $s_\theta(\cdot, t)$ is implemented. This is trained to approximate the score function $\nabla \log p_t(\cdot)$ such that, 
\begin{equation}
s_\theta(\cdot, t) \approx \nabla\log p_t(\cdot).
\end{equation}

The objective function for $s_\theta(\cdot, t)$ is the continuous weighted combination of Fisher divergences, given by, 
\begin{equation}
\mathbb{E}_{U(0,T)}\mathbb{E}_{p_{\mathbf{x}}(t)}[\lambda(t) \|\nabla \log p_t(\mathbf{x}) - s_\theta(\mathbf{x}, t)\|^2],
\end{equation}
where $U(0, T)$ denotes a uniform distribution over time interval $[0, T]$ and $\lambda$ is a positive weighting function to balance the magnitude of different score-matching losses.

Once the score-based model $s_\theta(\cdot, t)$ is trained, numerically solving the FBSDE is the same as the prediction of the form of a function (trajectory) that becomes the solution. In the literature, there are various methods to solve FBSDE, such as the Euler-Maruyama method and stochastic Runge-Kutta method. The key point here is that once a score predictor is obtained, then any form of SDE solver can be used to solve the reverse SDE numerically.

This paper particularly employs the Euler-Maruyama method to perform the backward SDE. This algorithm provides a method for denoising data by simulating the backward trajectory of the stochastic process. The Euler-Maruyama method is effective for integrating the backward SDE, provided the score function can be accurately estimated, which can be described as follows,
\begin{equation}
dX_t = [f(X_t, t) - g(t)^2 \nabla_x \log p_t(X_t)] dt + g(t) d\bar{W}_t,
\end{equation}
where $\nabla_x \log p_t(X_t)$ is the score, obtained by a neural network defined as $s_\theta(X_t, t) \approx \nabla\log p_t(X_t)$. 

For noise mitigation problem, the backward SDE process can be re-defined as follows,
\begin{multline}
    {d|\psi_t\rangle} = \left[f(\psi_t, t)\, - g^2(\psi_t,t)\nabla_\psi\log{p_t(|\psi_t\rangle)} \right]\,dt +
    \\ g(\psi_t,t) \,{d\bar{W_t}}.
\end{multline}

The backward process starts with $X_T$ sampled from a predefined noise distribution, typically Gaussian at the final noise level $T$.
The final noisy state $|\psi_T\rangle$ at time $T$ corresponds to $X_T$ in the noise mitigation problem, which is to be denoised eventually. 

The underlying dynamic in score-based diffusion models~\cite{song2020score} can be described by the Ornstein-Uhlenbeck diffusion process, i.e., 
\begin{align}
    dX_t &= -\alpha X_t \, dt + \sqrt{2}\beta \, dW_t, \\
    X_0 &\sim p_0, 
\end{align}
and $\alpha > 0$. 

The noising process $(X_t)_{t \in [0, T]}$ begins with data from $p_0$ and is perturbed into the invariant distribution $p_T (p_{\text{prior}})$. Generally, assumes $p_{\text{prior}}$ to be Gaussian, i.e., 
\begin{equation}
p_{\text{prior}} = N(0, 1/\alpha).
\end{equation}

When $ t $ approaches infinity, the distribution $X_t$ tends towards $ p_{\text{prior}}$, which is Gaussian (Ornstein-Uhlenbeck process). This can be seen as a situation where noise is added as time evolution progresses, which is defined in the dissipative term in (\ref{eq:LindbladlinearSDE}). The evolution $(|\psi_t\rangle)_{t \in [0, T]}$ starting from the noiseless state $|\psi_0\rangle$ in open quantum system problem to the noisy state $|\psi_T\rangle$.

The $(X_t)_{t \in [0, T]}$ dynamics reverses $p_t$ through time-reversal operations, using an explicit drift and diffusion matrix to precisely reverse the noise process, as follows,

\begin{align}
    dX_t &= [\alpha X_t \, dt + 2 \nabla \log p_{T-t} (X_t)]\, dt + \sqrt{2}\beta \, dW_t, \\
    X_0 & \sim p_T,
\end{align}
which can be formulated in our problem over the quantum computing domain, as follows,

\begin{align}
     {d|\psi_t\rangle} =& [\alpha |\psi_t\rangle \, dt + 2 \nabla \log p_{T-t} (|\psi_t\rangle)]\, dt \nonumber \\ 
     & + \sqrt{2}\beta \, dW_t, \\
    |\psi_0\rangle \sim& p_T.
\end{align}

The reconstructed data, $X_0$, represents an estimate of the original sample before noise was added. Moreover, $|\psi_0\rangle$ will be the noise-eliminated state in the noise mitigation problem.

\section{Concluding Remarks}\label{sec:Conclusion}
The primary aim of the paper is to mitigate errors in open quantum systems, a critical challenge as quantum computing transitions from theoretical exploration to practical application. This paper presents a new approach using diffusion models, specifically through the formulation of noise as forward-backward stochastic differential equations (FBSDE) and using the score-based generative model (SGM) for error mitigation. The proposed method leverages SGM to effectively denoise quantum states. This is achieved by dynamically adjusting the system's state towards an ideal, noise-free trajectory, thereby improving the precision and reliability of quantum computations. 

In conclusion, this paper not only advances our understanding of quantum error mitigation in open systems but also presents a practical and scalable approach that could significantly enhance the performance and reliability of quantum computers. This makes it a valuable contribution to the field of quantum computing, particularly in the era of NISQ.

\section*{Acknowledgment}
Joongheon Kim is a corresponding author.
\bibliographystyle{IEEEtran}
\bibliography{output}
\end{document}